\documentclass[proceedings]{JHEP3}

\conference{International Europhysics Conference on HEP}

\usepackage{epsfig,multicol}

\newcommand\fverb{\setbox\pippobox=\hbox\bgroup\verb}
\newcommand\fverbdo{\egroup\medskip\noindent%
            \fbox{\unhbox\pippobox}\ }
\newcommand\fverbit{\egroup\item[\fbox{\unhbox\pippobox}]}
\newbox\pippobox

\def\lesssim{\mathrel{\hbox{\rlap{\hbox{\lower4pt\hbox{$\sim$}}}\hbox{$<$}}}}
\def\gtrsim{\mathrel{\hbox{\rlap{\hbox{\lower4pt\hbox{$\sim$}}}\hbox{$>$}}}}
\newcommand{\ffrac}[2]{\left( \frac{#1}{#2} \right)}

\title{Upward and Horizontal $\tau$ Airshowers by UHE Neutrinos}

\author{\speaker{D.Fargion} \\

    Physics Department,INFN,Rome University 1,Italy\\
    E-mail: \email{daniele.fargion@roma1.infn.it}}
\received{\today}       
\revised{\today}
\accepted{\today}       

\preprint{\hepth{9912999}}  

\abstract{Upward and Horizontal $\tau$ Air-showers (UPTAUS and
HORTAUS) emerging from the Earth crust , mountain chains  or deep
plate boundaries are the most powerful signals of Ultra High
Energy UHE neutrinos $ \bar{\nu}_e $ at PeV and $\nu_{\tau}$,
$\bar\nu_{\tau}$ at energies  near and above $ 10^{15}-10^{19}
eV$. The large $\tau$  Air-showers multiplicity N in  secondaries
 $N_{opt} \simeq 10^{12} (E_{\tau} / PeV)$,
 $N_{\gamma} (< E_{\gamma} > \sim  10 \, MeV ) \simeq 10^8
(E_{\tau} / PeV)$, $N_{e^- e^+} \simeq 2 \cdot 10^7
(E_{\tau}/PeV)$, $N_{\mu} \simeq 3 \cdot 10^5
(E_{\tau}/PeV)^{0.85}$ make easy their discovery. UHE
$\nu_{\tau}$, $\nu_{\tau}$  because of neutrino flavor mixing,
($\nu_{\mu}\leftrightarrow \nu_{\tau}$),
 should be as abundant as $\nu_{\mu}$, $ \bar\nu_{\mu}$.
 Also $\bar{\nu}_e$, near the Glashow W resonance peak, $E_{\bar{\nu_e}} = M^2_W / 2m_e \simeq 6.3 \cdot
10^{15}\, eV$ may generate $\tau$ Air-showers. The HORTAUS may
test the UHE neutrino interactions  leading to additional
fine-tuned test of New TeV Physics  in  Mountain Valleys and
Earth crust  horizontal edges.  UPTAUS or HORTAUS, beaming toward
high mountains, air-planes, balloons and satellites should flash
 $\gamma$, $\mu$, X and Cherenkov lights toward detectors. Such
UPTAUS might already hit nearby most sensitive satellite as Gamma
Ray Observatory (GRO) detectors flashing them by short
(millisecond), hard, diluted $\gamma-$ burst at the edge of
threshold. We claimed their identity with the observed Burst And
Transient Source Experiment (BATSE) $78$ Terrestrial Gamma Flashes
(TGF). The TGF  clustering toward Galactic Center and Plane, known
galactic and extra-galactic sources strongly support their UHE
$\nu_{\tau}$ origin.} \keywords{UHE neutrino -$\tau$ air
shower-GZK} \dedicated{Dedicated to Giorgio Perlasca  heroic acts
in Budapest, 1944}
\begin{document}
 It is well known that Ultra High Energy $UHE$ neutrino of astrophysical origin above tens TeV
might overcome the nearby noise vertical of secondary atmospheric
neutrinos. The latter, being secondaries of charged cosmic rays,
smeared by galactic magnetic fields, have lost their interesting
astrophysical source records. In  cubic kilometer underground
detectors, both ice or water, in order to avoid the noisy downward
atmospheric muons and to overcome the Earth opacity for vertical
upward tens-TeV neutrinos, one \cite{01} should better neglect
vertical signals and focus the attention mainly on Horizontal
Underground detectors in kilometers wide disk-like or ring-like
arrays finalized to trace horizontal UHE Muons and Taus ($10^{13}-
10^{18}eV$) born by UHE astrophysical $\nu_{\mu}$,$\nu_{\tau}$.
Moreover because of $\tau$ amplified showering we prefer to
suggest the UPTAUS and HORTAUS detection  (after their parental
UHE $(\nu_{\tau},\bar{\nu}_{\tau}) + N$ interactions in rock, the
$\tau$ ejection in air and their fast decay in flight) as the best
tool for UHE neutrino discovery.
\FIGURE{\epsfig{file=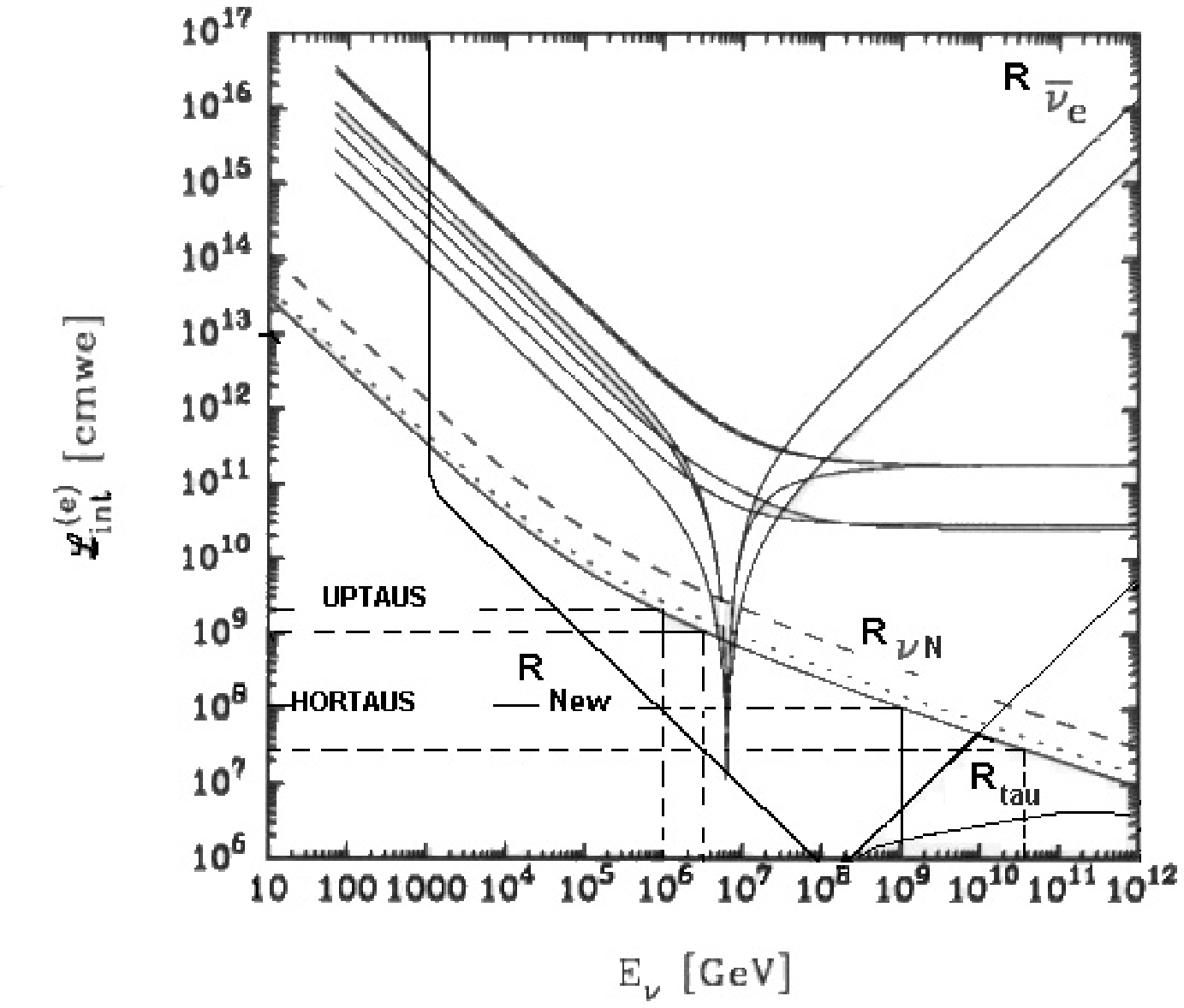,width=0.6\textwidth }
        \caption[]{The  UHE neutrino
    ranges as a function of the incoming UHE neutrino energy in Earth with
    overlapping the resonant $\bar{\nu}_e e$, $\nu_{\tau} N$ interactions;
    below  in the corner the growing UHE $\tau$ boosted Lorentz  range $R_{tau}$
     at the  same energies and in water lowest curve bounded by photo-pion interactions.
      Finally the solid line $R_{New}$  shows the interaction length due to New physics (extra dimension Gravity)
    at TeV for a matter density of rock $\rho =3$.\cite {01}, \cite {02}}%
    \label{Fig001}}
    Indeed UHE $\nu_{\tau}$ and $\bar{\nu_{\tau}}$ may be flavor
converted from common pion secondaries: $\nu_{\mu}$ and
$\bar{\nu_{\mu}}$. The UHE neutrinos $\bar{\nu_e}$,
${\nu}_{\mu}$, $\bar{\nu}_{\mu}$, are expected Ultra High Energy
Cosmic Rays  (UHECR) ( $\gtrsim 10^{16}$ eV) secondary products
near Active Galactic Nuclei (AGN) or micro-quasars jets by common
photo-pion decay relics by optical photons nearby the source,
either pulsars (PSR) or AGNs ($p + \gamma \rightarrow n + \pi^+,
\pi^+ \rightarrow \mu^+ \nu_{\mu}, \mu^+ \rightarrow e^+ \nu_e
\bar{\nu}_{\mu} $), or directly by proton proton scattering in
interstellar matter. UHE  neutrino  flavor  mix even at highest
Greisen-Zatsepin-Kuzmin (GZK) energy ($ > 4\cdot 10^{19} eV$)
because of the large galactic (Kpcs) and extreme cosmic (Mpcs)
distances much longer than oscillation ones:
 \begin{equation}
 L_{\nu_{\mu} - \nu_{\tau}} = 4 \cdot 10^{-3} \,pc \left(
 \frac{E_{\nu}}{10^{16}\,eV} \right) \cdot \left( \frac{\Delta m_{ij}^2
 }{(10^{-2} \,eV)^2} \right)^{-1}
 \end{equation}
  HORTAUS  are better detectable in deep valleys or on front
 of large mountain chains  like Alps, Rocky Mountains, Grand Canyons, Himalaya
  and Ande just near present AUGER project \cite{03},\cite{05}. Future Array
  Telescope may trace at best such EeV ($10^{18}$
 eV) HORTAUs in  the Death
 Valley in USA by photoluminescent tracks \cite{04}. The mountain chains screens undesirable  horizontal
$(>70^o)$ UHECR showers; HORTAUS may   lead also to UHE
horizontal muon bundle. The Mountain chains acts also as a
characteristic  $\bar{\nu_e}$ detector  at Glashow energy peak ($
6.3 PeV$). Present UPTAUS   is analogous to the  well-known
\cite{06}  " $\tau $ double bang". The novelty lays in the
explosive $\tau$ decay in air after its escape from the rock leading to amplified tau air-showers
in flight. The UPTAUS-HORTAUS channels  reflects the known  $\tau$ decay modes (Fig.2). 

\FIGURE{\epsfig{file=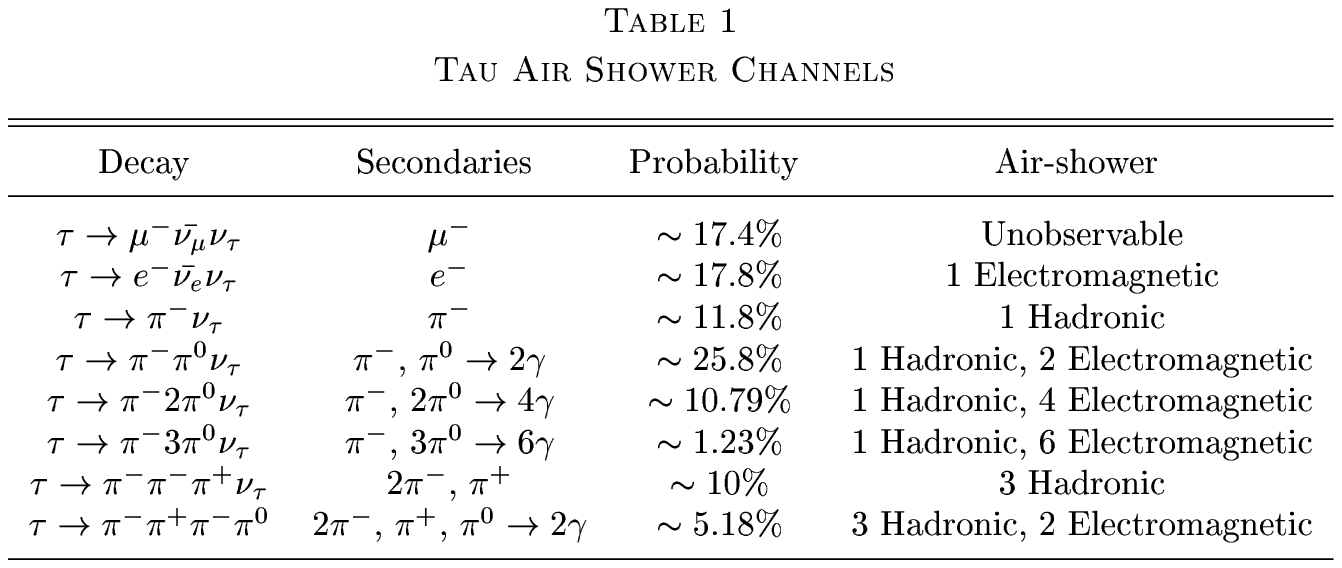,width=0.8\textwidth }
        \caption[]{The characteristic decay channels of UPTAUS and HORTAUS.}%
}

From the top of a mountain, a balloon or a satellite  the Earth
acts also as a huge target for UPTAUS or its wide corona crust for
HORTAUS. Observing from a height h downward toward the Earth at
any angle $\theta$ below the horizontal line,($\theta$ +$\pi/2$ =
zenith angle), the distances $d{(\theta)}$ toward the ground,
(from where an UPTAUS or HORTAUS should arise) is:
 \begin{equation}
 d{(\theta)}= (R_{\oplus}+ h) \cdot \sin\theta -
 \sqrt{(R_{\oplus}+ h)^{2}\cdot \sin^{2}\theta -(2hR_{\oplus} + h^2 )}
 \end{equation}
The distance length at horizontal tangential angle $\theta_{c}$
where the square-root term above vanishes, simplify in:
\begin{equation}
d{(\theta_{c})} = \sqrt{(2 R_{\oplus}\cdot h)+ h^2}\simeq
110\sqrt{ \frac{h}{km} }\cdot km
 \end{equation}
Where $\theta_{c} = \arcsin \sqrt{(2 h/ R_{\oplus})}\simeq 1.01
^\circ \sqrt{(h/km) }$; the terrestrial cord  distances
$\bigtriangleup d{(\theta)}$ crossed by the primary UHE
$\nu_{\tau}$ (and partially by the consequent upcoming $\tau$
before its exit in air) is
\begin{equation}
 \bigtriangleup d{(\theta)}= 2
  \sqrt{(R_{\oplus}+ h)^{2}\cdot \sin^{2}\theta -(2hR_{\oplus} + h^2 )}
 \end{equation}
 Such distances, which vanish for $\theta =\theta_{c}$, are not too long
 to suppress horizontal UHE neutrinos for small  $\delta\theta$ ($= \theta-\theta_c$) angles above $\theta_{c}$,
  even at energies $E_{GZK}$  $ \simeq 4 \cdot 10^{19} eV$ energies. see Fig 1. The  $\nu_{\tau}$+${\tau}$ crossed distances are $ \bigtriangleup
  d(\delta\theta)$:
\begin{equation}
 \bigtriangleup d{(\delta\theta)}\simeq 2(R_{\oplus}+ h)
  \sqrt{\delta\theta \sin 2\theta_c }\simeq  318 km \sqrt {\left(
 \frac{\delta\theta}{1^{\circ}} \right) \sqrt {\left(
 \frac{h}{km }\right)}}
 \end{equation}
 The  terrestrial surface below any high level observer covers huge areas
  A  ($ A = 2\pi R_{\oplus} (1-cos(\theta_c)) \simeq 2 \pi h  R_{\oplus} \simeq 4\cdot 10^4 \cdot km^2 (h/km)$ for $h <<
  R_{\oplus}$); however for too distant UPTAUS  origination the shower signal might be bounded by
   the longer crossed slant depth atmosphere opacity.  The effective  area for UPTAUS observed from height h $ > h_\circ$
  ($h_\circ$ is the atmosphere exponential length $\simeq 8.55 km$ ) is  smaller:
  $A_{eff} = \pi \cdot \cot^2(\theta) h^2 = 942 \cdot km^2 (h/ 10 \cdot km)^2; (\theta= 60^\circ).$
  The Tau decay track (see the line $R_{\tau}$ in Fig.1), constrained by  the characteristic distance horizons $d{(\theta_{c})}$
  defines a fine tuned HORTAU energy: $E_{\tau} = 2\cdot 10^{18} eV \sqrt{(h/km)}$. This formulas cannot be extended to arbitrary
  energy (or any height h), because of the finite atmosphere size; see
  below. A too large tau lifetime may lead to $\tau$ decay in too empty
  atmosphere.
 Keeping care of the Earth opacity, at large nadir angle ($\gtrsim {60}^0$)
where an average Earth density may be assumed ($< \rho > \sim 5$)
the transmission probability and creation of upward UHE $\tau$ is
approximately:
\begin{equation}
P(\theta,\, E_{\nu}) = e^{\frac{- \bigtriangleup
d{(\theta)}}{R_{\nu_{\tau}}(E_{\nu})}} (1 - e^{-
\frac{R_{\tau}(E_{\tau})}{R_{\nu_{\tau}}(E_{\nu})}}) \, .
\end{equation}
This expression should contain $\bigtriangleup d{(\theta)}$ from
above equation and the ranges $R_{\nu_{\tau}}$ and $R_{\tau}$
\cite{01}are shown in Fig 1; for example at PeV the  above
probability is within a fraction of a
million(${\theta}{\approx}{60}{^0}$) to a tenth of thousands
(${\theta}{\approx}{\theta_c}$). At GZK energies only HORTAUS are
allowed. The corresponding angular integral effective volume
observable from a high mountain (or balloon) at height $h$
(assuming a final target terrestrial density $\rho = 3$) for
UPTAUS at 3 PeV (for any  AGN neutrino flux model
  normalized within a  flat spectra whose
energy fluence $ \phi_{\nu}\simeq 2 \,10^{3} \frac{eV}{cm^{2}\cdot
s}$), is:
\begin{footnotesize}
\begin{equation}
  V_{eff} \approx 0.3 \, km^3 \ffrac{\rho}{3}\ffrac{h}{km} e^{-
  \ffrac{E}{3\,PeV}}
  \ffrac{E}{3\,PeV}^{1.363}
\ffrac {\phi_{\nu}}{2 \,10^{3}\frac{eV}{cm^{2}\cdot s}}
\end{equation}
\end{footnotesize}
 Any  AGN neutrino flux model
  normalized within such a  flat spectra
is leading, above 3 PeV, to $\sim$ 10 UHE $\nu_{\tau}$ upward
event/km$^3$ year \cite{02}. The consequent average upward
effective event rate  on a top of a mountain (h $\sim 2\,km$) is:
\begin{footnotesize}
\begin{equation}
  N_{eff} \simeq 8 \, \frac{\mathrm{events}}{\mathrm{year}} \ffrac{\rho}{3} \ffrac{h}{2 \,km}
e^{- \ffrac{E}{3\,PeV}} \ffrac{E}{3\,PeV}^{1.363} \ffrac
{\phi_{\nu}}{2 \,10^{3}\frac{eV}{cm^{2}\cdot s}}
\end{equation}
\end{footnotesize}
Their signals at ten kms distances should be $\phi_{\gamma} \simeq
10^{-4} \div 10^{-5} cm^{-2} s^{-1}$, $(\phi_{X \sim 10^5 eV})
\simeq 10^{-2} \div 10^{-3}cm^{-2} s^{-1})$. The optical
Cherenkov flux is large $\Phi_{opt} \approx 1 cm^{-2}$.
We claimed \cite {01} that such UPTAUS or HORTAUS produce gamma
bursts at the edge of GRO-BATSE  originated from the Earth and
named consequently as Terrestrial Gamma Flashes (TGF).
The effective volume for UPTAUS and the event rate within an angle
of view ($\theta> 60^o$) is, at 3 PeV,  approximately to within 15
km$^3$ values and the expected UHE PeV  rate is:
\begin{equation}
  N_{ev} \sim 150 \cdot e^{-
  \ffrac{E_\tau}{3\,PeV}}
  \ffrac{E_\tau}{3\,PeV}^{1.363}
  \ffrac{h}{400 km} \,
  \ffrac {\phi_{\nu}}{2 \,10^{3}\frac{eV}{cm^{2}\cdot s}}
  \frac{\mathrm{events}}{\mathrm{year}}
\end{equation}
The TGF signals would be mainly $\gamma$ at flux $10^{-2}$
cm$^{-2}$ at X hundred keV energies.
 The correlations of these clustered TGFs directions
 toward   GeV-MeV (EGRET), X  sources, Milky Way Galactic Plane (Fig.3)  support
 and make suggestive the TGF  identification as UPTAUS and HORTAUS. The TGF location reflects the higher UPTAUS (and HORTAUS)
 interaction probability in the rock  over the sea (and along the
coastal plates). Highest magnetic field on Asia  widely spreads
UPTAUS  making the TGF more observable. The present TGF-$\tau$
could not  be produced by UHE $\bar{\nu}_e$  because of the
severe Earth opacity and support the $\nu_{\mu}\leftrightarrow
\nu_{\tau}$  flavor mixing.
     The new physics interaction at TeV while forbid upward UHE signals in
     underground $km^3$ detectors it will amplify the $\nu_{\tau}$ signals
     beyond mountains,  by two order of magnitude making extremely fruit-full
     UHE $\nu_{\tau}$ astrophysics in near future.
\FIGURE{\epsfig{file=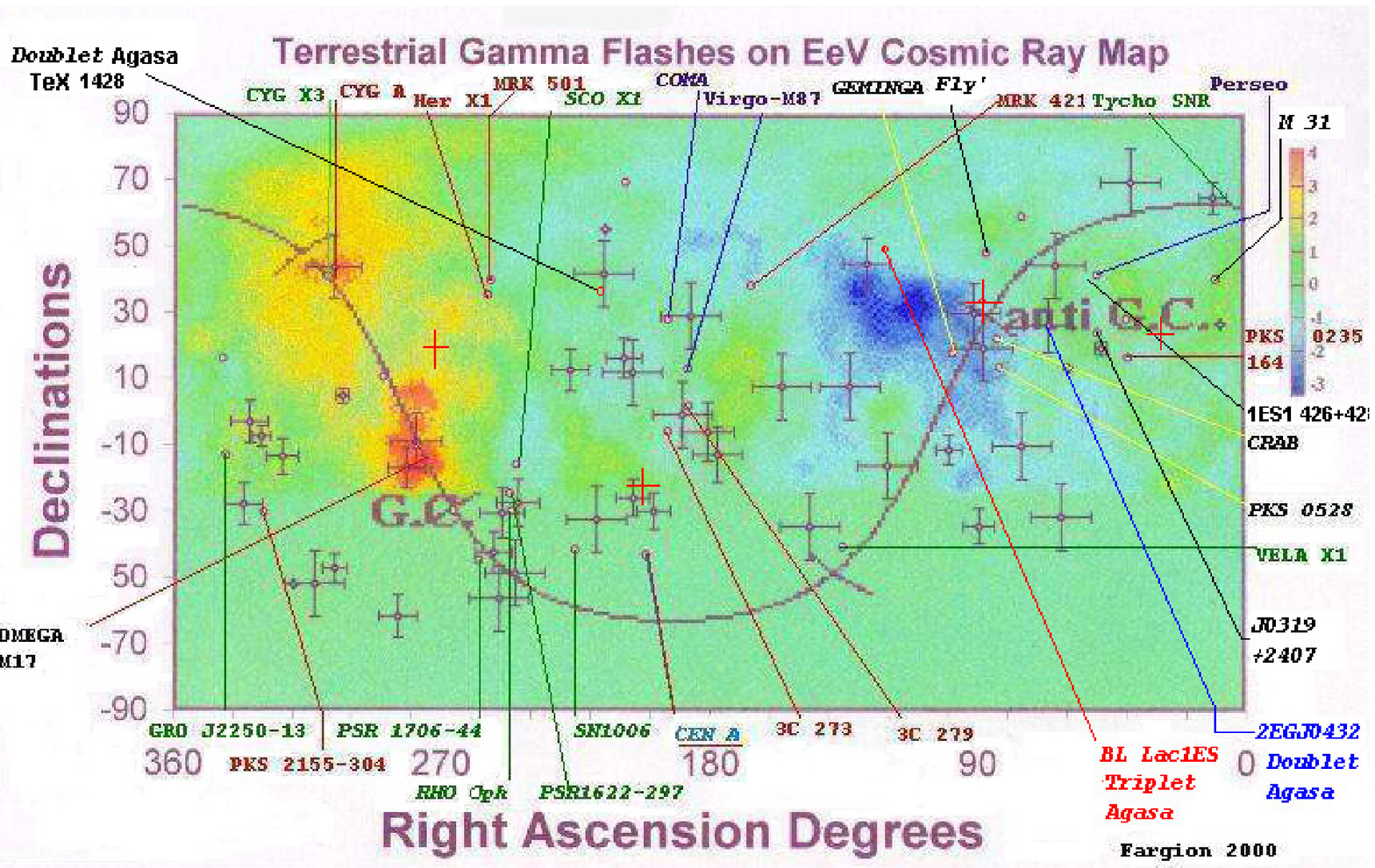,width=0.8\textwidth }
        \caption[]{ TGF events toward galactic center, disk and known
 sources on EeV AGASA map. Four red cross mark last TGF events. \cite{01}.}%
    \label{Fig003}}
  HORTAUS may develop nor at too dense low atmosphere (being absorbed), neither at too high, low atmosphere
(where no shower may be amplified). HORTAUS charged secondaries
may also turn upward by geo-magnetic fields into fan-thin- cone
jets, appearing as UPTAUS. The maximal $c\tau$ distance is ruled
by:
\begin{equation}\label{13}
  \int_{0}^{+ \infty} n_0 e^{-\frac{\sqrt{(c\tau+x)^2+R_\oplus^2} - R_\oplus}{h_0}}
   dx \cong  \int_{0}^{+ \infty} n_0 e^{-\frac{(c\tau+x)^2} {2h_0R_\oplus}}
   dx \cong n_0 h_0 A
\end{equation}
\begin{equation}\label{15}
 c\tau = \sqrt{2R_\oplus h_0}
 \sqrt{ln \ffrac{R_\oplus}{c\tau} - ln A }
\end{equation}
Where $A=A_{Had.}$ or $A=A_{\gamma}$ are parameters of order of
unity, logarithmic function of energy,  that calibrate the energy
shower slant depth for both hadronic or electro-magnetic
nature,\cite{04}: $A_{Had.}=0.792 \left[1+0.02523
 \ln\ffrac{E}{10^{19}eV}\right];$
  $A_{\gamma}=\left[1+0.04343\ln\ffrac{E}{10^{19}eV}\right].$
The solution of this equation leads to a characteristic UHE
$c\tau_{\tau}$ = $546 \;km$ decay distance  at height $h= 23$ km
where the HORTAUS start to shower. This imply a possibility to
discover efficiently by satellite and balloons arrays  UHE
$\nu_{\tau}$, $\bar\nu_{\tau}$ up to $1.11$ $10^{19}eV $.
\cite{04,07}. From high satellite  the arrival HORTAUS angle maybe
confused ($\mp  1^\circ$) with most common Albedo Horizontal High
Altitude Showers (HIAS) \cite{04}. However from balloons heights
and below, HIAS  arrival angles split ($\geq 7^\circ$) from
HORTAUS ones and are well  distinguished. There is also the
simplest possibility to observe UPTAUS and HORTAUS while they are
hitting and lightening, via Cherenkov lights, upward mountain
snow-walls. Such UPTAUS may also beam on lower boundary of high
altitude clouds in the nights. These reflected flashing lights
have a characteristic twin beam eight-shaped imprint that offers
to Telescopes a new kind of signature for UHE Neutrino
Astrophysics.



\end{document}